\documentclass[pmlr]{jmlr}% new name PMLR (Proceedings of Machine Learning)

\usepackage{longtable}% for long tables
\usepackage{booktabs} % for professional tables
\usepackage{hyperref}

 \usepackage{booktabs}
 
\usepackage[load-configurations=version-1]{siunitx} % newer version

\makeatletter
\def\set@curr@file#1{\def\@curr@file{#1}} %temp workaround for 2019 latex release
\makeatother

\theorembodyfont{\upshape}
\theoremheaderfont{\scshape}
\theorempostheader{:}
\theoremsep{\newline}

\jmlrvolume{[VOLUME \# TBD]}
\jmlryear{2022}
\jmlrworkshop{Under Review}

\title[Short Title]{Interaction of A Priori Anatomic Knowledge with Self-Supervised Contrastive Learning in Cardiac Magnetic Resonance Imaging}

\author{\Name{Makiya Nakashima}
       \Email{nakashm2@ccf.org}\\ 
       \addr Heart Vascular and Thoracic Institute\\
       Cleveland Clinic\\
       Cleveland, OH, United States of America 
       \AND
       \Name{Inyeop Jang}
       \Email{jang.inyeop@mayo.edu}\\ 
       \addr Mayo Clinic Cancer Center\\
       Mayo Clinic\\
       Jacksonville, FL, United States of America 
       \AND
       \Name{Ramesh Basnet}
       \Email{ramesh.basnet@circlecvi.com}\\ 
       \addr Circle Cardiovascular Imaging\\
       Calgary, Alberta, Canada 
       \AND
       \Name{Mitchel Benovoy}
       \Email{mitchel.benovoy@circlecvi.com}\\ 
       \addr Circle Cardiovascular Imaging\\
       Calgary, Alberta, Canada 
       \AND
       \Name{W.H. Wilson Tang}
       \Email{tangw@ccf.org}\\ 
       \addr Heart Vascular and Thoracic Institute\\
       Cleveland Clinic\\
       Cleveland, OH, United States of America 
       \AND
       \Name{Christopher Nguyen}
       \Email{nguyenc6@ccf.org}\\ 
       \addr Heart Vascular and Thoracic Institute\\
       Cleveland Clinic\\
       Cleveland, OH, United States of America 
       \AND
       \Name{Deborah Kwon}
       \Email{kwond@ccf.org}\\ 
       \addr Heart Vascular and Thoracic Institute\\
       Cleveland Clinic\\
       Cleveland, OH, United States of America 
       \AND
       \Name{Tae Hyun Hwang}
       \Email{hwang.taehyun@ccf.org}\\ 
       \addr Mayo Clinic Cancer Center\\
       Mayo Clinic\\
       Jacksonville, FL, United States of America 
       \AND
       \Name{David Chen}
       \Email{chend3@ccf.org}\\ 
       \addr Heart Vascular and Thoracic Institute\\
       Cleveland Clinic\\
       Cleveland, OH, United States of America} 

\editor{Editor's name}

\begin{document}

\maketitle

\begin{abstract}
Training deep learning models on cardiac magnetic resonance imaging (CMR) can be a challenge due to the small amount of expert generated labels and inherent complexity of data source. Self-supervised contrastive learning (SSCL) has recently been shown to boost performance in several medical imaging tasks. However, it is unclear how much the pre-trained representation reflects the primary organ of interest compared to spurious surrounding tissue. In this work, we evaluate the optimal method of incorporating prior knowledge of anatomy into a SSCL training paradigm. Specifically, we evaluate using a segmentation network to explicitly local the heart in CMR images, followed by SSCL pretraining in multiple diagnostic tasks. We find that using a priori knowledge of anatomy can greatly improve the downstream diagnostic performance. Furthermore, SSCL pre-training with in-domain data generally improved downstream performance and more human-like saliency compared to end-to-end training and ImageNet pre-trained networks. However, introducing anatomic knowledge to pre-training generally does not have significant impact. 
\end{abstract}

\section{Introduction}
One of the driving elements for the wide spread success of deep learning (DL) is the availability of large amounts of labeled data. Much of the work where DL models achieved comparable accuracy and sometimes even surpassing human physicians \cite{RN620, RN93, RN69, RN290}, required tens to hundred of thousand annotated cases. Such volume of labeled data is only feasible for large healthcare institutions which have access to both the human resources to annotate exams and see the volume of patients needed to generate the raw data. Even then, many of the most urgent medical uncertainties cannot hope to accumulate the necessary data given the long-tail of disease phenotyping \cite{RN957}.
	The lack of data volume and expert annotation resources are further complicated by the nature of cardiac imaging. Cardiac imaging often entails capturing short video clips of the beating heart in multiple different views. Cardiac magnetic resonance imaging (CMR) further adds to the data complexity by capturing different image contrasts which are sensitive to different cardiac physiologies and pathologies. This flexibility makes CMR a very powerful imaging modality; but also one that is highly time-consuming and expertise-demanding to interpret correctly as cardiac diseases can present in different locations and different contrasts. Computationally, the variability of CMR images which incorporates both multi-contrast 2D images and short videos of the heart, makes designing a self-supervision task specific for CMR difficult. 
	Training DL models naively using small amounts of data often do not achieve desirable accuracy for clinical usage \cite{RN7}. One solution is to leverage pre-trained models from other domains through transfer learning. For example, models pre-trained on natural images in ImageNet \cite{RN120} are widely used for medical tasks \cite{RN958,RN948}. However, medical images have very different image characteristics and properties compared to natural images, which sometimes results in poor generalizability \cite{RN958} or may even produce worse results \cite{RN948}. 
	Self-supervision learning (SSL) offers a data or label efficient solution to pre-train a network using in-domain data. Unlike supervised or semi-supervised learning, SSL utilizes tasks (e.g. image denoising \cite{RN952}, inpainting \cite{RN745}, jigsaw puzzles \cite{RN955}, etc.) which implicitly teaches a model to learn representations of objects, actions, or context in the data domain. The learned representation can then be transferred to downstream tasks of interest. Another option to reduce the dependency on labeled data is to introduce domain knowledge to constrain the search space of networks \cite{RN1113}. 
	Yet is it unclear how generalizable these methods are, particularly when applied to such a complex data domain such as CMR. Furthermore, although CMR is heavily focused on interrogating diseases affecting the heart, the images contain adjacent organs which are often sources of spurious features. In this work, we propose constraining the trained representation to only the organ of interest in the context of SSL. 

\subsection*{Generalizable Insights about Machine Learning in the Context of Healthcare}
- Representations trained using natural images such as ImageNet do not produce optimal representations for CMR\\
- A priori knowledge of salient areas can greatly improve model performance with small amounts of data\\
- Contrastive self-supervision is a powerful pretraining tool which minimizes need for annotated data\\
- A priori knowledge does not always improve self-supervised pretrained models and needs to be evaluated on a case-by-case basis

\section{Related Work}

More efficiently using in-domain data is an important area of research in medical imaging analysis. We classify methods into two broad categories: unsupervised in-domain pretraining and leveraging domain specific knowledge. 

\subsection{Unsupervised in-domain pretraining}
Given that it is widely known that networks trained from in domain (e.g. medical images) generalize better than networks trained on out of domain (e.g. natural images) data. One popular method of pre-training involves applying deformations to images and training and model to recover the original image. Specific tasks proposed for medical images include image denoising \cite{RN952}, inpainting \cite{RN745}, and solving jigsaw puzzles \cite{RN955}. Chen et al \cite{RN695} presented an inpainting method computed tomography and MRI which involved randomly shuffling small patches in the image. The model would then need to rely on contextual clues to recover the original image. Similarly, one could also pose contextual learning by splitting an image into patches and mixing it up similar to solving a jigsaw puzzle \cite{RN700}. The model then tasked to identify the order of the original image. However, this work was evaluated on images of static organs. Working with multi-view CMR images, Bai et al leveraged the natural intersection between intersecting 2D images in a 3D space to learn anatomic landmarks \cite{RN949} although found quickly diminishing returns in even moderately sized datasets. These pre-training tasks can also be used in conjunction with each other, as Koohbanani et al \cite{RN698} found a combination of different tasks including magnification prediction, jigsaw, and discriminating fake images produced better results than individual tasks in pathology images.

Self-supervised contrastive learning (SSCL) has recently been proposed as an alternative in SSL paradigms whereby the model is trained to maximize the similarity within the data embeddings themselves rather than any specific outcomes or tasks. This property enables SSCL to learn embeddings which are robust to common image deformations such as changes in brightness, rotation, and random cropping compared to individual tasks. Truong et al \cite{RN904} found SSCL consistently outperformed supervised medical image classification models trained from scratch in digital pathology, fundus imaging, and X-ray images. SSCL has also been shown to augment global embeddings of CMR images trained using image reconstruction techniques by additionally maximizing similarities in local context \cite{RN950}. However, this work looked only into image segmentation tasks which is a small portion of medical workflow. 

\subsection{Leveraging domain specific knowledge}
Adding a priori knowledge implicitly to either the training process or explicitly to the model itself is another strategy to minimize data needs and improve model performance. Xie et al \cite{RN1113} classifies training methods into 3 categories: 1) training the models in human-like manner, 2) explicitly defining hierarchical diagnostic patterns, and 3) pre-defining saliency. Overall, many of these methods involve splitting the training process into multiple, easier tasks. For example, Wang et al leveraged a hierarchical model to identify areas of abnormality in the general field of view of chest X-rays and then provide disease specific diagnosis using locally identified patches \cite{RN1114}. Identifying abnormal areas is an easier task than producing a specific diagnosis while using localized patches for diagnosis can artificially increase the volume of training data. Yet such a much works best if the disease is localized. Mitsuhara et al more directly used a combination of direct classification and semantic segmentation of disease to improve automatic grading of retinal images \cite{RN1115}. However, this strategy works best if the disease is well characterized by morphologic abnormalities rather than physiological ones. 

Furthermore, very few of these methods have demonstrated improvements in the inherent representation of the data. CMR is a highly complex imaging modality where both morphology and texture are important for diagnosis and prognosis. And although the standard views captured in CMR generate fairly consistent views of the heart, the extra-cardiac tissue captured in the standard field of view can have huge variation which may contribute to difficulties creating a robust representation of the data. Therefore, it is important to understand how limiting field of view to only the heart may change the performance of deep learning networks for CMR applications. 

\section{Methods}
\subsection{Anatomic segmentation}
We automatically identify the salient cardiac structure using cvi42 (Circle Cardiovascular Imaging, Calgary, Canada) automatic cardiac segmentation algorithm. The method identifies short axis cine slices and automatically generates segments of the left ventricular blood pool, left ventricular myocardium, and one comprehensive right ventricular segment for all time frames. We combine these segments to produce a general cardiac mask for short axis cine time frames although we primarily use the end systolic and end diastolic frames. An experienced clinician then manually quality controlled a portion of the masks. 

\subsection{simCLR}
We learn visual representation of CMR images using SSCL through simCLR \cite{RN837}. Contrastive learning creates a deep representation of data by maximizing agreement between images from the dataset and augmented versions (a positive pair) while minimizing agreement between two random images and their augmentations (negative pair). This training forces the network to ignore features associated with simple image augmentations such as cropping, Gaussian noise, contrast jitter, and rotation and focus on learning features which are inherent to the image domain (e.g. shape of the myocardium). The network used to encode image features $h_i$=$f(x_i)$ can be any chosen network ranging from simple multi-layer CNN to DenseNets to vision transformers. We can then project the encoder with a projection head $g(h_i)$ to reduce the potential loss of information induced by the contrastive loss. We use the normalized temperature-scaled cross entropy loss as a measure of positive pair similarities as given below:

\begin{equation}
    \resizebox{.5\linewidth}{!}{$
    \displaystyle
    \mathcal{L}_{i,j}=-log \frac{exp(cos(z_i,\hat{z}_i)/\tau)}{\sum_{k=1}^{2N}\mathbb{1}_{k\neq i}exp(cos(z_i,\hat{z}_i)/\tau)}
$}
\end{equation}

Where the cosine similarities of the projected features of the positive pair is normalized by the sum of the similarities between all negative pairs. $\tau$ is the temperature hyperparameter which controls local separation and global uniformity. The total loss in each mini-batch is the sum of the loss from all positive pairs.

\section{Cohort}
We evaluated the self-supervision approach on three different CMR datasets and two different clinical tasks to identify the generalizability of self-supervision methods. The first dataset is the open source ACDC \cite{RN743} data of short-axis “cine” CMRs. The cine images refer to high resolution, high contrast images which capture a short clip of physiologic motion. The video clips generally contain between 16-30 frames and cover a single cardiac cycle. The ACDC dataset includes multi-class labels for normal (NOR), dilated cardiomyopathy (DCM), hypertrophic cardiomyopathy (HCM), prior myocardial infarction (MINF), and abnormal right ventricle (RV) as well as labels for semantic segmentation of cardiac anatomy. The other two datasets are organization specific CMR datasets of patients with cardiomyopathies including non-ischemic cardiomyopathy (NICM), ischemic cardiomyopathy (ICM), amyloid (AMYL), and HCM. These datasets represent a variety of cardiac diseases, diagnostic tasks, and cleaniness of data.
\subsection*{ACDC}
The ACDC dataset \cite{RN743} is a popular open source CMR cine dataset. The dataset comprises of 150 clinical CMR acquired at the University Hospital of Dijon, France acquired over a 6 year period. The images were acquired on a combination of 1.5T Siemens Area and 3.0T Siemens Trio scanner. The dataset provides only short axis views, in-plane spatial resolution ranging from 1.37 to 1.68 mm\textsuperscript{2}, and slice thickness of 5-8mm. The images cover 90-100\% of the cardiac cycle with 28-40 images. Given that these are clinical scans, there is a variety of noise levels, artifacts, field-of-views, and LV coverage. The dataset includes two sets of labels/tasks; balanced multi-class disease classification and semantic segmentation of cardiac anatomy. The dataset covers 4 cardiac diseases (myocardial infarction with systolic heart failure, dialated cardiomyopathy, hypertrophic cardiomyopathy, and abnormal right ventricle) and “normal” classified according to the medical reports. All instances of disease are considered unambigious with clear delineation between.
\subsection*{Internal datasets (ICM/NICM and HCM/AMYL)}
Two retrospective datasets were constructed of adult patients who underwent a CMR exam between 2008 and 2020. One dataset (ICM/NICM) includes 272 patients diagnosed with ischemic cardiomyopathy (ICM) and 733 patients diagnosed with non-ischemic cardiomyopathy (NICM). The 2nd dataset (AMYL/HCM) consists of 194 patients diagnosed with cardiac amyloidosis (AMYL) and 260 patients diagnosed with hypertrophic cardiomyopathy (HCM). All patients were referred to our institution for suspected cardiomyopathy. The patients received a standard CMR exam, with cine, and late gadolinium enhancement (LGE) imaging on a Phillips 1.5T Achieva or 3T Ingenia scanners. The data also includes a wide variety of pulse sequence parameters. All diagnoses were determined by a level 3 expert CMR cardiologist. Trained clinical annotators then went through clinical records for diagnosis, imaging study findings, and clinical biomarkers. Annotators also segmented a subset of 227 NICM cine images for left ventricular myocardium, left ventricular cavity, and right ventricle using CMR42 (Circle Cardiovascular Imaging, Calgary, Canada). A board certified cardiologist with 10+ years experience reading CMR reviewed the segments for quality assurance.
	Usage of these datasets for research purposes was approved by the Institutional review board and granted with waiver for consent.
\begin{table}[b]
  \centering 
  \caption{AUCs of models trained using the full, non-masked images. Bolded results indicate the best performing training paradigm for that specific dataset/model combination. None refers to no pretraining.}
  \label{full} 
  \begin{tabular}{lccccc}
  \toprule
%    \textbf{Dataset} & \textbf{Model} & \textbf{None} & \textbf{ImageNet} & \textbf{Full-SSCL} & \thead{\textbf{Segmented} \\ \textbf{-SSCL}}\\
    Dataset & Model & None & ImageNet & Full-SSCL & Segmented-SSCL\\
    \midrule
    ACDC & VGG16 & 0.542 & 0.532 & \textbf{0.849} & 0.738\\ 
    ACDC & DenseNet121 & 0.574 & 0.766 & \textbf{0.867} & 0.765\\ 
    ICM & VGG16 & 0.724 & \textbf{0.877} & \textbf{0.877} & 0.860\\ 
    ICM & DenseNet121 & 0.843 & 0.901 & \textbf{0.902} & \textbf{0.902}\\ 
    AMYL & VGG16 & 0.570 & 0.605 & 0.684 & \textbf{0.698}\\ 
    AMYL & DenseNet121 & 0.621 & 0.667 & \textbf{0.708} & 0.698\\ 
    \bottomrule
  \end{tabular}
\end{table}
\section{Results on Real Data} 
\subsection{Study Design and Evaluation}
We evaluate the effect of using an anatomic prior to pretrain deep representation of short axis cine CMR with SSCL. We compare the SSCL pretrained representations with representations pretrained on ImageNet and no pretraining at all. Since VGG16 and DenseNet121 are developed primarily for 2D data, we used a separate model for the end diastolic and end systolic frames. We then concatenated the final layer together to achieve a single model using both frames.
	We split each dataset into train and test datasets, a randomized 70/30 split for each dataset. Optimal hyperparameters were found through cross validation only on the training data. We used early stopping to prevent the model from overfitting during the pre-training phase. We then fine-tuned a fully connected output layer for classification. For pre-training, we did a grid search on the batch size from 64-256. We kept the temperature and learning rate at 0.1 and 1e-4 respectively. We used random crop, contrast, vertical and horizontal flipping, and rotation for wdata augmentation. For the training phase, all models were trained using Adam optimizer with a weight decay of 1e-4. We used sparse categorical cross-entropy loss. We then test the models using the test sets associated in each fold. Because the classes were relatively balanced, we evaluated model performance for classification tasks using macro area under the receiver operating curve (AUC). 
\subsection{Classification Accuracy in Different Datasets}
\begin{table}[t]
  \centering 
  \caption{AUCs of models trained using segmented images. Bolded results indicate the best performing training paradigm for that specific dataset/model combination. None refers to no pretraining.}
  \label{seg} 
  \begin{tabular}{lccccc}
  \toprule
%    \textbf{Dataset} & \textbf{Model} & \textbf{None} & \textbf{ImageNet} & \textbf{Full-SSCL} & \thead{\textbf{Segmented} \\ \textbf{-SSCL}}\\
    Dataset & Model & None & ImageNet & Full-SSCL & Segmented-SSCL\\
    \midrule
    ACDC & VGG16 & 0.746 & 0.814 & 0890 & \textbf{0.901} \\ 
    ACDC & DenseNet121 & 0.744 & \textbf{0.890} & 0.888 & 0.878 \\ 
    ICM & VGG16 & 0.791 & 0.826 & 0.886 & \textbf{0.908} \\ 
    ICM & DenseNet121 & 0.856 & 0.892 & 0.886 & \textbf{0.904} \\ 
    AMYL & VGG16 & 0.594 & 0.625 & \textbf{0.680} & 0.647 \\ 
    AMYL & DenseNet121 & 0.597 & 0.637 & 0.669 & \textbf{0.730} \\ 
    \bottomrule
  \end{tabular}
\end{table}
Table 1 show the results of pretraining models using the full, non-segmented image. It is clear from the consistently poor performance of randomly initialized models that each of these datasets are not well suited to fully train a network end-to-end. It is no surprise given the low volume relative to data complexity. Models trained in this manner regularly vastly underfit the data compared to any sort of pretraining, particularly in the ACDC dataset where each class has only 14 unique cases. Training models using random initialization on the internal datasets produce in more competitive results, possibly due to both the higher volume and fewer number of classes although far more variation in image characteristics. In this case, masking the images during pretraining did not improve results compared to using the full image to pretrain the models. This is possibly due to the additive value of the surround tissue. 
	In contrast, when the training data was itself segmented (Table 2), the models that were pretrained using segmented data performed better. However, the most significant effect were to models which were not pretrained or used ImageNet pretraining. Here, the results improved significantly particularly in the low-data ACDC dataset. This would suggest the high additive value of using anatomic prior knowledge to constrain the search space. In comparison with using the full image (Table 3), we observe that using segmented images in all phases of training (supervised training and self-supervised pretraining) still generally produced the best results.
\begin{table}[h]
  \centering 
  \caption{AUCs comparing pretraining using different combination of segmentation. Bolded results indicate the best performing training paradigm for that specific dataset/model combination.}
  \label{comp} 
  \begin{tabular}{lccccc}
  \toprule
    Dataset & Model & Full-Full & Full-Segmented & Segmented-Full & Segmented-Segmented\\
    \midrule
    ACDC & VGG16 & 0.849 & 0.890 & 0.738 & \textbf{0.901} \\ 
    ACDC & DenseNet121 & 0.867 & \textbf{0.888} & 0.765 & 0.878 \\ 
    ICM & VGG16 & 0.877 & 0.886 & 0.860 & \textbf{0.908} \\ 
    ICM & DenseNet121 & 0.902 & 0.866 & 0.902 & \textbf{0.904} \\ 
    AMYL & VGG16 & 0.684 & 0.680 & \textbf{0.698} & 0.647 \\ 
    AMYL & DenseNet121 & 0.708 & 0.669 & 0.698 & \textbf{0.730} \\ 
    \bottomrule
  \end{tabular}
\end{table}

\subsection{Pre-training Reduces Convergence Time}

In figure 1, we show the rate of convergence of the pre-trained models in the internal AMYL/HCM dataset. We find that all the SSCL pretrained models converge faster than either the ImageNet pretrained models or the not pretrained models. Using segmented images does not seem to impact the rate of convergence but interestingly, seems to cause more instability in the models. 

\begin{figure}[t]
  \centering 
  \includegraphics[width=6in]{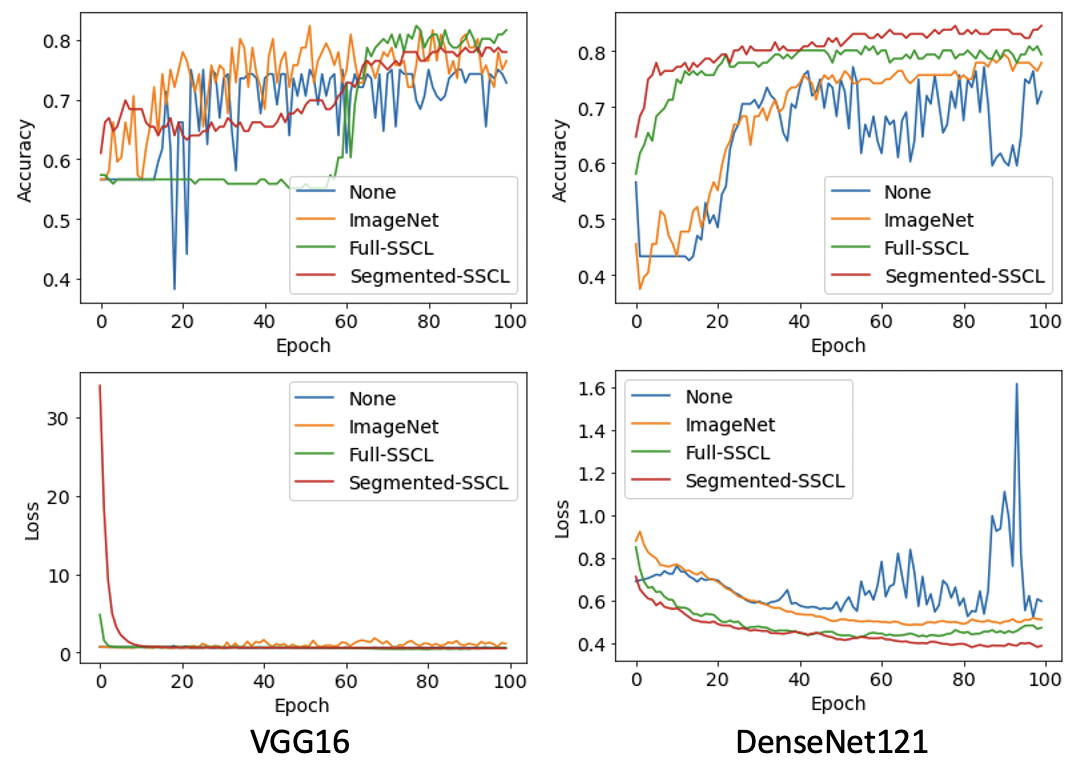} 
  \caption{Convergence of pre-trained models for downstream classification. }
  \label{converge} 
\end{figure}

\section{Discussion} 
In this work, we explore the impact of using anatomic priors alongside modern SSCL pretraining to constrain clinical predictive models using low volume annotated CMR data. Removing potential superfluous tissue by segmenting the heart can greatly improve predictive power even using very small amounts of data. Adding this a priori knowledge of anatomy can also work synergistically with modern pretraining methods such as SSCL, generating extremely good results. However, it should also be noted that segmentation methods can introduce errors in the image and may inadvertently remove informative areas. This is especially important in the context that SSCL can generate similar results without the need to train a segmentation model. 

Although SSCL does not require any labels and therefore can leverage large amounts of unlabeled data, SSCL can also be extremely time consuming to train as the representations often improve with larger batch sizes and/or longer training time \cite{RN837}. Chen et al found that batch sizes of up to 4096 was beneficial for SSCL. These batch sizes demand cutting edge hardware and/or long training times which is not readily available to all groups. On the contrary, training a segmentation model can be fairly computationally efficient and can be done with relatively small amounts of data \cite{RN545}. Incorporating such a priori knowledge can be a very computationally efficient way to make large gains in prediction accuracy as shown in Table 2. 

These findings are significant for many clinical applications as most clinical datasets are small in number. Relying on naïve fully supervised training will often yield poor results. Identifying the most suitable priors and training methodologies can be massive impact on achievable training accuracy and therefore, potential downstream impact on patients. 

\paragraph{Limitations}

This work explores only contrastive learning in a specific subset of CMR data; short axis cine images. It is currently unknown how important anatomic priors are to the full range of CMR data, which is necessary for the diagnosis of a wide range of cardiac diseases. In future work, we will explore how anatomic priors and SSCL can be applied with different views (e.g. long axis, 3 chamber, etc.) and different contrasts (e.g. T1 weighted, T2 weighted, late gadolinium enhanced). However, it is clear that both anatomic priors and modern self-supervised pretraining techniques can greatly improve model performance, particularly in the low data regime. 

\paragraph{Data Availability}
The ACDC dataset is freely available at https://www.creatis.insa-lyon.fr/Challenge/acdc/databases.html. Internal institutional datasets utilized in this study are not publicly available due to privacy and security concerns. The data is not easily redistributable to researchers other than those engaged in Institutional Review Board-approached research collaborations with this intution.

\paragraph{Code Availability} 
All code used will be made freely available at upon publication at https://github.com/placeholder.

% ACKNOWLEDGEMENTS ONLY GO IN THE CAMERA-READY, NOT THE SUBMISSION
% \acks{Many thanks to all collaborators and funders!}
\bibliography{ssl_cmr_v1.5}

\end{document}